\documentclass{mem}
\usepackage{natbib}\usepackage{txfonts}\usepackage{balance}
\usepackage{graphicx}
\usepackage[a4paper]{hyperref}
\idline{75}{282}
\begin{document}
\def\teff{$T\rm_{eff }$}
\def\kms{$\mathrm {km s}^{-1}$}
\newcommand{\lsp}{LS~I~+61$^{\circ}$303}
\newcommand{\grs}{GRS 1915+105~}
\newcommand{\grp}{GRS 1915+105}
\newcommand{\lsi}{LS~I~+61$^{\circ}$303~}
\newcommand{\ls}{LS~5039}
\title{
Steady jets and transient jets:
observational  characteristics and models
}

   \subtitle{}

\author{
M. \,Massi\inst{} }

% \offprints{M. Massi}

\institute{
Max Planck Institut f\"r Radioastronomie,
Auf dem H\"ugel 69,
D-53121 Massi, Germany
\email{mmassi@mpifr-bonn.mpg.de}
}

\authorrunning{Massi }

\titlerunning{Steady jets and transient jets}

\abstract{
Two types of radio emission are observed from X-ray binaries with jets. They have
completely different
characteristics and are associated with different kinds of ejections. 
One kind of emission  has  a  flat or inverted
spectrum 
indicating  optically thick self-absorbed
synchrotron emission;
the second kind of emission corresponds to an optically thin ``transient'' outburst.
The flat or inverted spectrum covers the whole radio
band and has been established also at millimeter and infrared wavelengths. When this
kind of radio
emission is spatially resolved it appears as a continuous jet, the so-called
``steady jet''.
In contrast, transient jets associated with optically thin events are 
resolved as   ``plasmoids'' moving at relativistic speeds away from the center of the system.
The most important point is that the two kinds of radio emission and their
corresponding types of ejections
seem to be  related to each other;  the optically thin outburst that characterizes
the transient jet 
occurs  after an interval of emission with  flat/inverted  spectrum. 
The hypothesis that the two  classes of  ejections correspond to two different 
physical processes is corroborated by  X-ray observations.
The transient jet is associated with the steep power-law X-ray state,
whereas the continuous jet always corresponds to
the  low/hard  X-ray state.
Two different models successfully describe the two jets: a conical flow and shocks.
The  conical outflow describes  the continuous  jet and
internal shocks in a  continuous pre-existing outflow describe  the  ``plasmoids'' of
the transient jet.
The internal shocks in the outflow are thought to originate 
from a new population of very fast  particles.
Three open issues are discussed: 
is magnetic reconnection  
the physical process generating the new population of 
very fast  particles?
Is that part of the  continuous jet  called  ``core''
destroyed by the transient jet and its associated shocks?
Can we extrapolate these results from steady and transient jets in X-ray binaries
to radio loud AGNs? 
}
\maketitle{}

\section{Introduction}

In the past, Seyferts galaxies, Quasars and radiogalaxies 
have been thought to be  
quite different objects, 
because of  the quite different  features  observed in  each of them. 
But later, it became clear that all of them are  members of the same class, 
the active galactic nuclei (AGN) \citep{antonucci93},
%(Urry and Padovani 1995),
that is: supermassive black holes accreting from the host galaxy.
Whereas the majority of AGNs are weak in their radio emission, about 10\% 
are hundreds to thousands times stronger and are called ``radio-loud''
\citep{barvainis05}.
The radio emission in radio-loud AGN  originates from a jet.
%Richard Barvainis et al. 2005 10% 

\begin{figure}
\resizebox{\hsize}{!}{\includegraphics[clip=true]{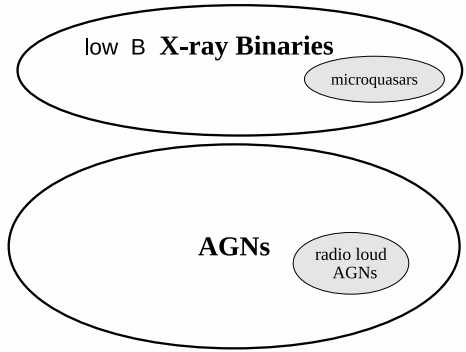}}
\caption{\footnotesize Radio-emitting relativistic jets
 are observed in both
 classes of accreting compact objects: 
Active galactic nuclei (AGN) and low B X-ray Binaries 
(see text).}
\label{f1}
\end{figure}

X-ray binary systems are formed by a compact object and a normal  star.
X-ray binaries where the compact object is either
a stellar mass black hole or a neutron star with 
a low magnetic field 
($\leq 3 \times 10^8$G, \citet{massi08})
are the galactic equivalent to AGNs: 
the compact object accretes from the companion star and 
in some cases a radio emitting jet is observed 
 (Fig. \ref{f1}).
The first X-ray binary with an associated radio jet was SS433 \citep{spencer79};
%(Spencer 1979); 
in the 90's several radio-emitting X-ray binaries where discovered and 
named    ``microquasars'' \citep{mirabel93}.

With the seminal papers
of \citet{dhawan00},
  \citet{fender01} and \citet{fender04}
%(2000; 2002; 2004 ?) 
it became clear that there are two different jet types: a  relatively  steady, continuous jet and a   ``transient''
jet.
The two types of jets have  different  spectral and morphological radio 
characteristics. In addition, when observed in X-rays the microquasars
show different spectral X-ray states depending on the jet type.

Here I review  these two type of jets,  their different observational
characteristics in radio 
and X-rays (Sec. 2) and their models (Sec. 3). 
A number of  open questions are commented in Sec. 4.
\section{Observational characteristics}
\begin{figure*}
\resizebox{\hsize}{!}{\includegraphics[clip=true]{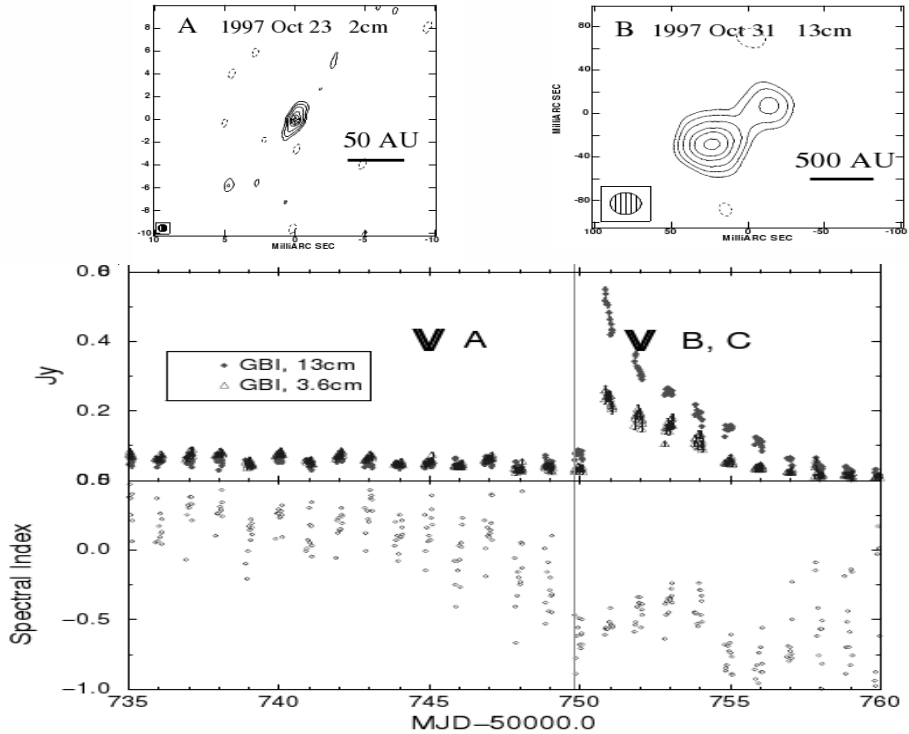}}
\caption{\footnotesize Steady jet and transient jet: spectral index characteristics
and morphology \citep{dhawan00}}. 
%Dhawan, Mirabel and Rodriguez 2000. The spectral  index (bottom curve) }
\end{figure*}
In this section I compile different observational aspects of systems with 
steady or  transient jets in the radio band and in X-rays.
From radio observations it results that 
there seem to be  differences in the spectral index ($\alpha$), 
the morphology, the velocity and in polarization. 
X-ray differences are mainly indicated by the photon index ($\Gamma$) of the power
law fitted to the spectrum.
%and astrometry. 
%\subsection{Spectra}
\subsection{Radio properties}

%{\it Spectral index}

The spectral index $\alpha$, 
defined as 
 $\alpha={log(S_1/S_2)\over log(\nu_1/\nu_2)}$,
with  flux density  $S\propto \nu^{\alpha}$, is 
 $\geq 0$ 
in  steady jets  
(indicating a flat or inverted spectrum) 
whereas in transient jets   $\alpha <$0
(see  Fig. 4 in \citet{fender01}.
The  flat or inverted spectrum,  $0\leq \alpha \leq 0.6$,
covers the whole radio band 
and  has been established also at  millimeter and infrared wavelengths 
\citep{fender00, russell06}.
%Fender et al. 2000; Fender 2001; Russel et al. 2006).
For transient jets the spectral index results in
$-1\leq \alpha \leq -0.2$ \citep{fender01},
which corresponds to an index p=$1.4-3$ of the power-law energy distribution of the
relativistic electrons
responsible for the radio synchrotron radiation.
  %Also for AGn alpha between 0.6-0.7??? chiamano flat spectrum altro 
When  the  radio emission corresponding to a flat/inverted spectrum 
is observed  at high resolution  it appears as a continuous  jet
with a bright ``core'' \citep{dhawan00}. A transient jet shows up with an optically thin radio outburst.
Multifrequency radio observations of  GRO J1655−40 
show that during the   radio outburst 
the flux densities at all observing frequencies peak simultaneously, with the amplitude
of the flare increasing toward lower frequency 
\citep{hannikainen06}. 
%\citep{hannikainen06, stevens03}. 
                   %(Hannikainen 2006).
At high resolution 
the emission is resolved in  components,
some time called  plasmoids,
movig  apart \citep{mirabel94,fender99}.  
     %Vedi i due picchi di tarscher..
      %VELOCITA: per cygnus x-1 0.6

The steady jet re-establishes  quite soon after the optically thin outburst 
related to  the transient jet:
\citet{dhawan00}  observed in \grs  that the steady jet is  re-established  within 18 hrs from
the start of a major optically thin outburst. 
{\it This implies that  fast travelling  components  with optically thin spectrum
from the transient jet  may still dominate the radio emission,
but in reality they are detached from the actual situation around the engine
where the steady jet re-established.} 
%Among microquasars in one or in the other state, 
%{\it Polarization}
\citet{fender04} argue that the velocities of the transient jets are significantly
larger than those of the steady jets.
Finally, there is evidence that the radio emission from 
ejected plasmoids is stronger polarized than the emission from the continuous
jets \citep{fenderbelloni04}.
%Anche lis peracaula

Cygnus X-1 and \grs are the
best examples of systems switching between the two  kinds of ejections
(see references in Gallo 2009).
In Fig. 2,  we see the transition from continuous to transient jet
in spectra and morphology observed by 
\citet{dhawan00} for GRS 1915+105.
In Fig. 2, one also sees that the spectrum remains flat/inverted 
for $\sim~14$ days and then  
$\alpha$ switches to negative values.
The outburst is  optically thin, with both frequencies peaking simultaneously and
with the peak at the lowest frequency being more than a factor 2 higher than that at the higher
frequency.
In Fig. 2-top,  simultaneous VLBA images show the continuous jet (left) 
and  the transient jet (right).
From the asymmetry between approaching and receding components of the steady 
jet of  GRS1915+105, \citet{dhawan00}
%Dhawan and collaborators (2000) 
derived a mildly relativistic speed of $\beta=0.1$. 
From the proper motion  of plasmoids travelling in opposite directions from the
center a high value of $\beta \geq 0.9$ was derived 
\citep{mirabelrodriguez99}. 
% and equal to 0.98 in \citet{fender99}.  Altre velocita'.  Lsi taylor slow noi con
% asimmetrie 0.3c

%(Fender \& Belloni 2004).
%\subsection{The transient jet follows the steady jet}
\begin{figure}[]
\begin{center}
\includegraphics[scale=0.3, angle=-90.]{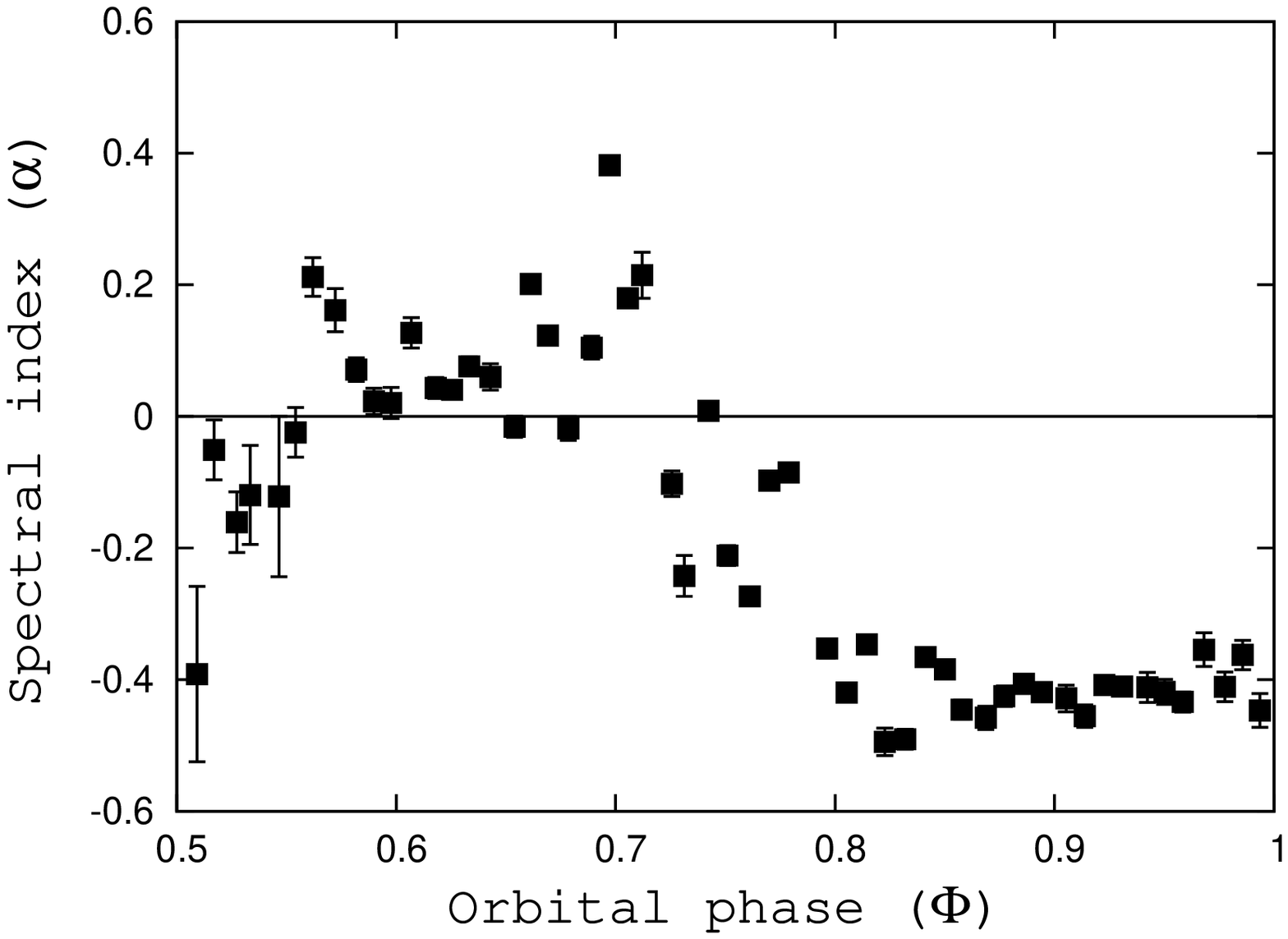}\\
\includegraphics[scale=0.3, angle=-90.]{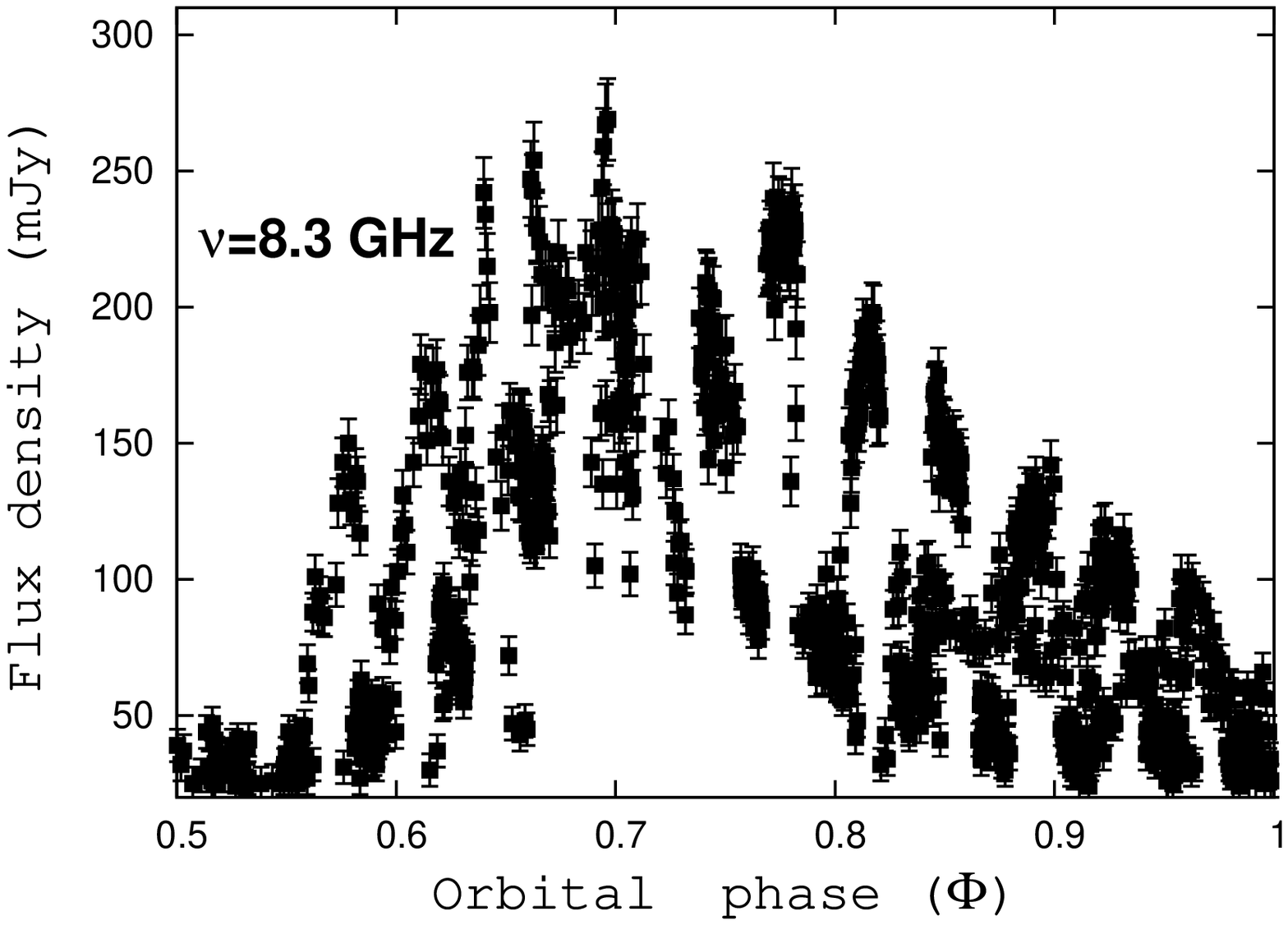}\\
\includegraphics[scale=0.3, angle=-90.]{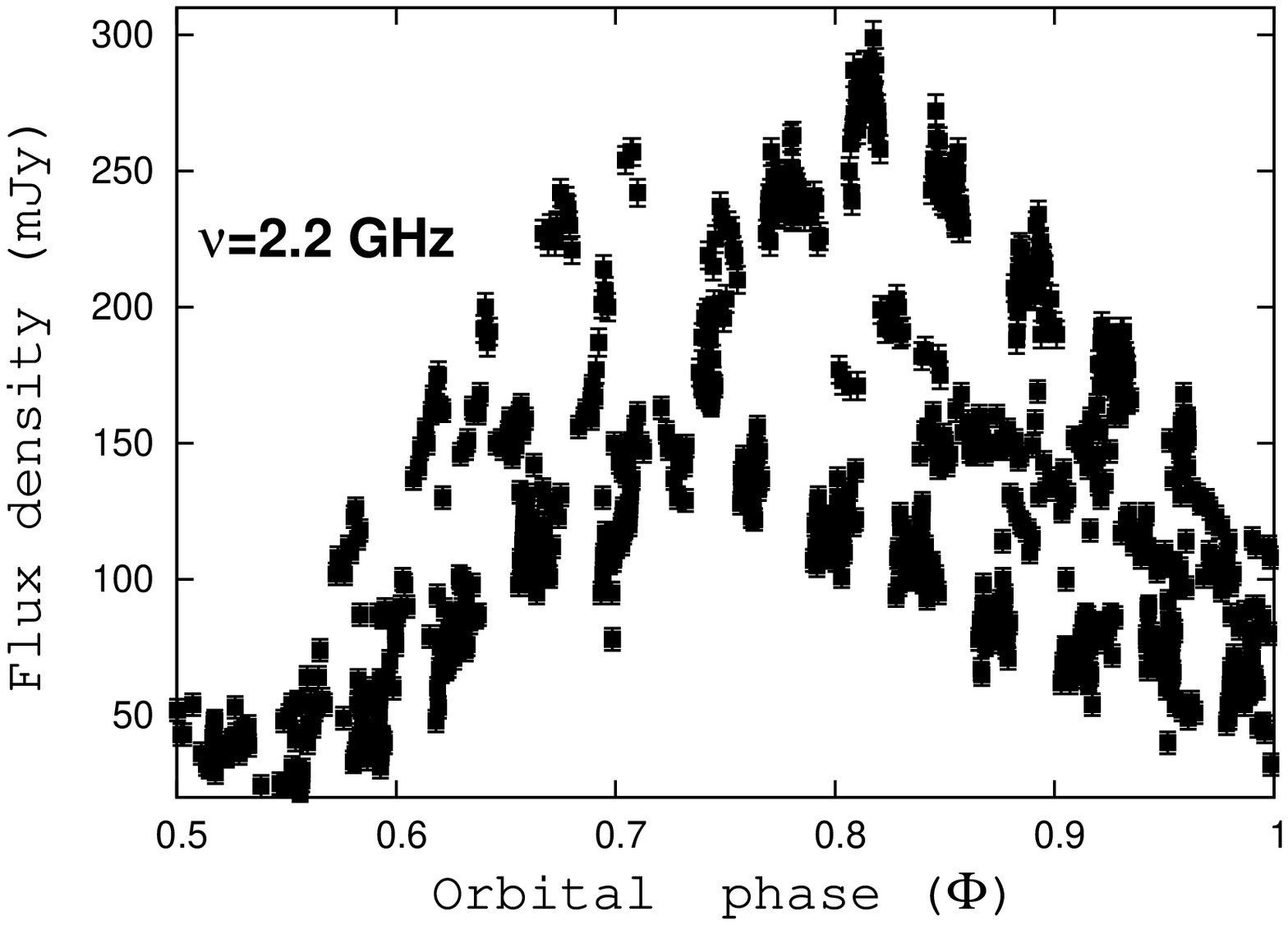}\\
\end{center}
\caption{\footnotesize
Spectral index and flux density, at 8.3 GHz and 2.2 GHz  vs orbital phase,
 $\Phi$ ($P_{orbit}$=26.5 d), 
for  Green Bank Interferometer  data of the periodical source \lsi
\citet{massikaufman09}).
The   radio ``outburst'' is in reality formed by two consecutive outbursts: the first
one optically thick (i.e. peaking at 8.3 GHz) and the second, stronger, optically thin
(i.e. peaking at  2.2 GHz).
Between the two outbursts there is
an   inversion of the spectrum from inverted to optically thin. 
}
\label{oo4}
\end{figure}

In \grp,  the continuous  jet with optically thick emission 
corresponds to a prolongued state of relatively high, but rather stable radio emission 
that in fact is called ``plateau''. 
In  the system \lsi, the optically thick radio emission phase corresponds to 
an increasing  emission level, related to an increasing accretion rate, 
terminating in an optically thick outburst
\citep{boshramon06, massi10}. 
After this inverted spectrum phase
the optically thin outburst occurs  
as in \grp.
Figure~\ref{oo4} shows,  folded with the orbital period of 26.5 d, 
light curves  of \lsi at 2.2 GHz and 8.3 GHz;  Figure~\ref{oo4}-top 
showes the  spectral index \citep{massikaufman09}.
At 8.3 GHz, we see  a main peak of $S_{8.3{\rm GHz}}=270 \pm 15$  mJy at $\Phi=0.69$.
We call this outburst  {\it Peak$_1$} (indicated by the vertical bar at $\Phi=0.69$ in Fig. 3), which is an optically thick
outburst, as one can determine from the clear positive spectral index  shown in the
figure (Top).
At 2.2 GHz, the figure shows that {\it Peak$_1$} decays, but then  the
flux increases again till a large  outburst at $\Phi=0.82$; we call this {\it Peak$_2$} (indicated by the vertical bar at $\Phi=0.82$ in Fig. 3),
($S_{2.2{\rm GHz}}=299 \pm 6$ mJy).  At 8.3 GHz {\it Peak$_2$} corresponds to a minor 
outburst resulting in a clear optically thin spectrum.
\begin{figure}[]
\begin{center}
\includegraphics[scale=0.3, angle=-90.]{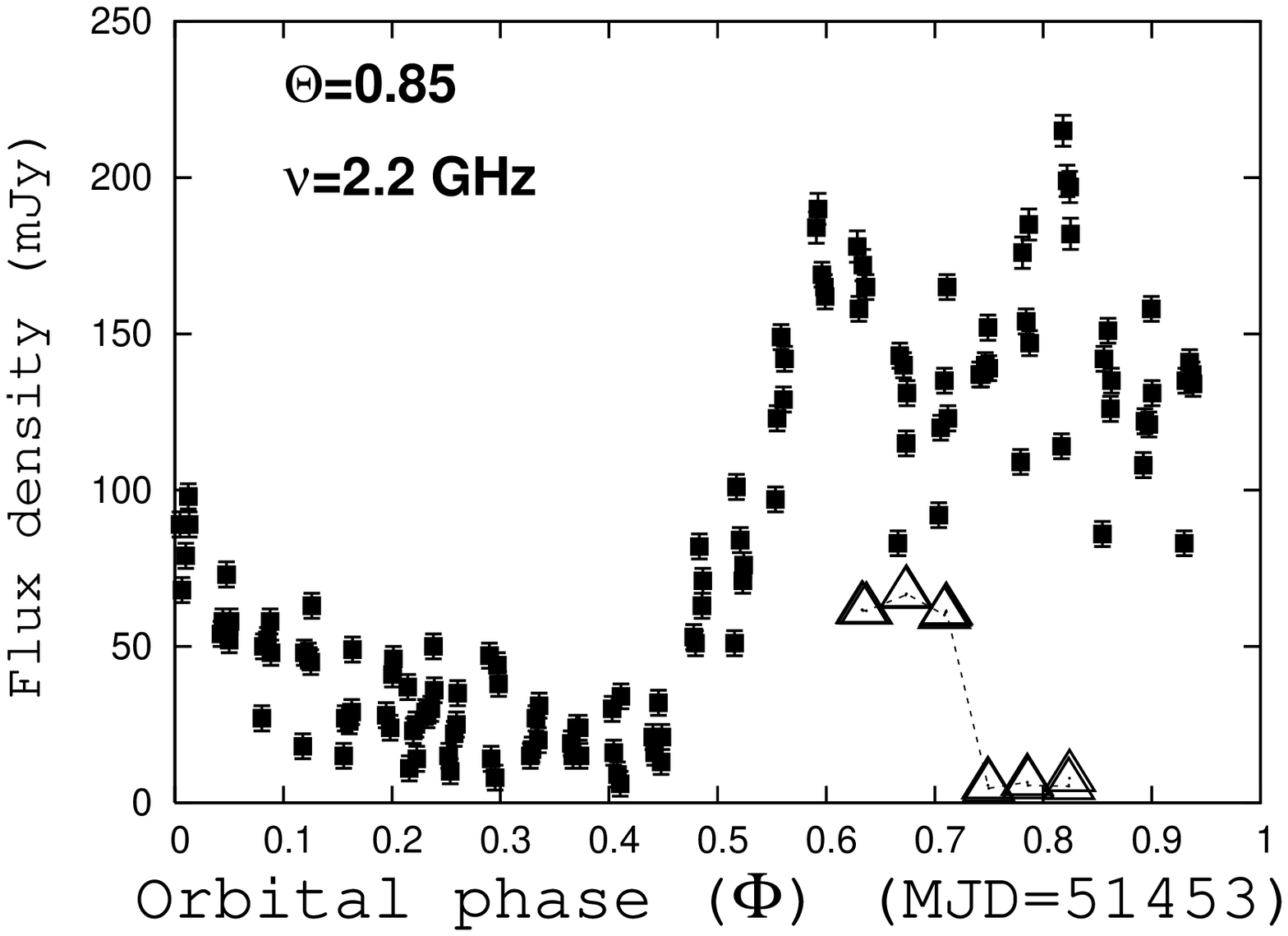}
\includegraphics[scale=0.3, angle=-90.]{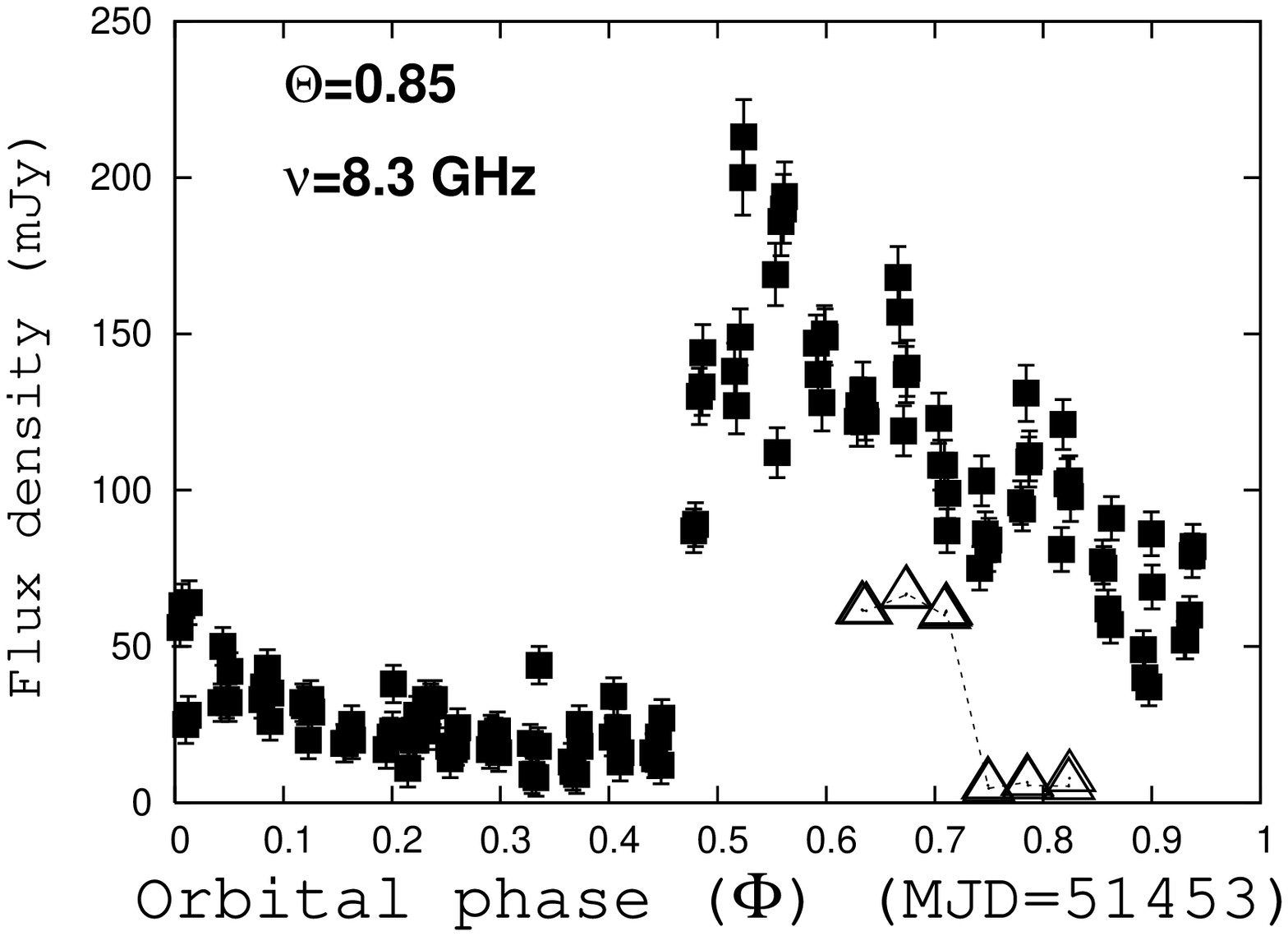}
\end{center}
\caption{\footnotesize Light curves of \lsi along the orbital phase  
at 2.2 GHz and 8.3 GHz \citep{massikaufman09}.
The triangles are the   H$\alpha$ emission-line 
measurements by \citet{ grundstrom07}
multiplied here by a factor of -5 to fit in the plot.} 
\label{oo5}
\end{figure}

In Fig.~\ref{oo5} we analyze an individual
 light curve (a 26~d data set).
{\it Peak$_1$} and {\it Peak$_2$} at 2.2 GHz are well distinguishable.
{\it Peak$_1$} at  2.2 GHz (Fig.~\ref{oo5}-Top) has a delay 
%of about $\Delta \Phi\simeq 0.1=2.6$ days 
with respect to {\it Peak$_1$}
at 8.3 GHz (Fig.~\ref{oo5}-Bottom), a behaviour consistent with 
an adiabatically expanding  cloud \citep{vanderlaan66}. Following this model, at this point an optically thin decay should follow.
On the contrary at $\Phi=0.8$ an optically thin outburst occurs.
The optically thin outburst,  {\it Peak$_2$}, 
%occurs,   within the uncertainty of the sampling ($\Delta \Phi=0.01??$) 
%at both frequencies simultaneously at $\Phi\simeq ??$..  
%{\it Peak$_2$} 
reaches $\sim$220 mJy at 2.2 GHz, whereas at 8.2 GHz it is only $\sim$130 mJy.
In the same plot (Fig.~\ref{oo5}-Bottom), are given the  H$\alpha$ emission-line
measurements by \citet{grundstrom07}.
The H$\alpha$ excess  present until  $\Phi=0.71$,
shows a dramatic decline at $\Phi=0.749$ (1 day later)
corresponding to the onset of the optically thin outburst.

The H$\alpha$ emission line observations  corroborate the scenario of 
two distinctly different kinds of  jets, corresponding to two different underlying physical processes.
Still the two jets must be related to each other: the transient jet occurs after the
steady jet. We will see in Sect. 3.2 how in fact the shock-in-jet model assumes for the
transient jet  a pre-existing (to the transient) slow flow (i.e. the steady,
continuous  jet).
\subsection{X-ray  properties}
  \citet{fender04}
  %(2000; 2002; 2004 ?)
pointed out, how the  
  two different types of radio jets correspond to two different
  X-ray spectral states.
When the microquasar showes a steady jet, 
its X-ray spectral state is the  low/hard state 
corresponding to  a power-law
with photon index $\Gamma \sim 1.7$ (2-20 Kev)
\citep{mcclintockremillard06}.
The optically thin radio outburst (i.e. the transient jet) is always associated  with  the
Very High State (VHS)\citep{fender04}.
The  power-law  characteristic for  this state  has  a photon index 
$\Gamma \sim 2.5$ and extends into the gamma-ray regime.  The VHS
was  renamed  steep power-law X-ray state, because  monitoring
programs of ${\it RXTE}$  showed that an unbroken steep power-law
is the fundamental property of the state \citep{mcclintockremillard06}. 
Figure 5 resumes the characteristics of these  X-ray states.
\citet{fender04} stated  that it may be exactly the point of the transition to the
steep power-law state
that corresponds to the optically thin radio outburst.
\begin{figure}
\resizebox{\hsize}{!}{\includegraphics[clip=true]{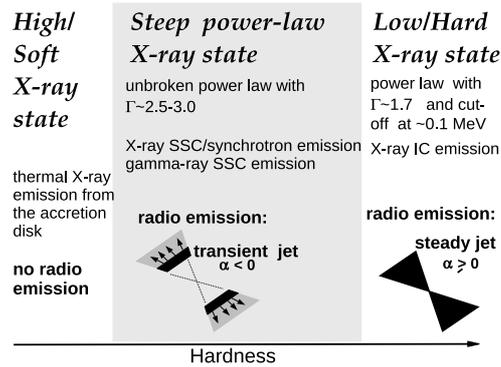}}
\caption{X-ray spectral states and related radio states 
(see \citet{massizimmermann10}).}
\end{figure}
\section{Models}
In this section, I briefly review  the two 
models that  describe the two jets:
a conical outflow for the continuous  jet and
internal shocks in a  continuous flow for the transient jet.
\subsection{Conical jet}
The formation of the steady  jets is  thought to be mediated by  large-scale open
magnetic fields
threading the rapidly rotating accretion disks
\citep{blandfordpayne82}.
% (i.e. magneto-centrifugal acceleration,  Blandford &amp; Payne, 1982).
A magnetized plasma containing energetic electrons with a power law energy distribution
will produce a synchrotron power-law spectrum.  However, below a critical frequency
($\nu_{break}$),
the radiating electrons will re-absorb some of the photons and as a result the
typical  spectrum
of a uniform synchrotron source shows a peak at $\nu_{break}$.
Changes in the electron energy distribution and the
decay of the magnetic field
along a conical jet imply that the critical  frequency varies along the jet.
Assuming several jet segments
(Fig. 6, see also Fig. 1 in \citet{marscher95}),
 each producing a spectrum with
a different $\nu_{break}$,   the composite spectrum will
appear flat (Kaiser 2006).
In microquasars, the  $\nu_{break}$ for the part of the steady jet closest to the engine (i.e.
 $\nu_{break_1}$) is in the infrared (Russel et al. 2010 and references there).
In AGNs
the observed turnover is at $10^{11\pm 0.5}$ Hz, called ``millimeter-wave core'' in
the literature
\citep{marscher95}.
Imaging a steady jet, gives rise to the effect known as ``core shift'', with the shift as a
function of the observing frequency  $\nu_{obs}$.
At $\nu_{obs}$ the emission of the segment will dominate, whose spectrum peaks at
that frequency
plus small contributions from neighboring segments (see Fig.1 in \citet{markoff10}).
\citet{dalymarscher88} calculated 
that  changes due to the external pressure  lead 
the jet
boundary to oscillate as the jet gas periodically
overexpands and reconverges in its attempt
to match the ambient pressure. 
This  effect of the boundary creates  
a network of waves in the interior of the jet that,  
converging  toward the axis,  may 
form a standing shock.
%\citet{dalymarscher88} calculated 
%that  changes due to the external pressure  lead to the formation within the jet of
%waves that reflecting inside the symmetric structure may converge to 
%form a standing shock.
Very recently,  \citet{jorstad10}
established  that the millimeter-wave core in AGNs
is a physical feature of the jet, i.e. coincident with
the standing shock,  that is different from the observed cm-wave core, the location
of which is determined by the above discussed jet opacity.

Models of the low/hard X-ray state for X-ray binaries
give  a geometrically thin, optically thick accretion disk truncated
at  $R_{tr}\simeq 100 r_g$ \citep{mcclintockremillard06,done10}.
Following Meier (2005), the terminal velocity of the  steady  jet
is approximately equal to the escape speed at the footpoint of the magnetic field
where the jet is launched. For the large  $R_{tr}\simeq$100 $r_g$ the escape velocity
drops to values below 0.6~c  in agreement  with the low velocities
inferred for steady jets. When a system is in a very low low/hard state, the even 
larger truncated radius
may result in a very low velocity
% \citep{massizimmermann10}.
(e.g. $0.06\pm 0.01$ for \lsp, \citet{peracaula98,massizimmermann10}).
The steady jet model therefore assumes  that the flat/inverted
spectrum could be the result of the variation of the magnetic field and the density of relativistic particles along a conical jet;
the  velocity  is  low  for jets  anchored at the large truncated radii associated to  
the low/hard state.

\subsection {The shock-in-jet model}

\citet{fender04}
associate the  change in the radio spectrum 
- from optically thick to optically thin -
to the parallel change that is observed in the X-ray states of these sources, when
passing from
the low/hard X-ray state to the steep power-law state. 
Fender and collaborators  make the hypothesis  that 
in such a passage 
there is  
an  increase in the bulk Lorentz factor of the ejected material.
{\it This increase gives rise to  shocks where
 the new  highly-relativistic plasma  catches
up with the pre-existing slower-moving material of the steady jet}.
This model, the  shock-in-jet model,  was originally  derived
by \citet{marschergear85} for AGNs, then generalized by \citet{valtaoja92,turler00},
and
introduced in the context of X-ray binaries for \grs by \citet{kaiser00} 
(see the review by \citet{turler10}).

In the  internal shock  model, variations of the jet velocity or pressure lead to
the formation of shock waves 
and electrons are accelerated.
% within a layer behind the shock front.
All frequencies will peak simultaneously. 
However, the higher energy emission particles die out first and
the highest frequency emission remains confined to a thin layer 
behind the shock front (Fig. 7) \citep{marscher10}.
The width of the  emission layer behind the shock front, $x$,
is inversely proportional to the square root of the radiated frequency. 
As a result the flare will dominate at  lower frequencies 
resulting in the observed  optically thin outburst.
%Ha a che fare con Amplitude o widht del picco????
\begin{figure}
\resizebox{\hsize}{!}{\includegraphics[clip=true]{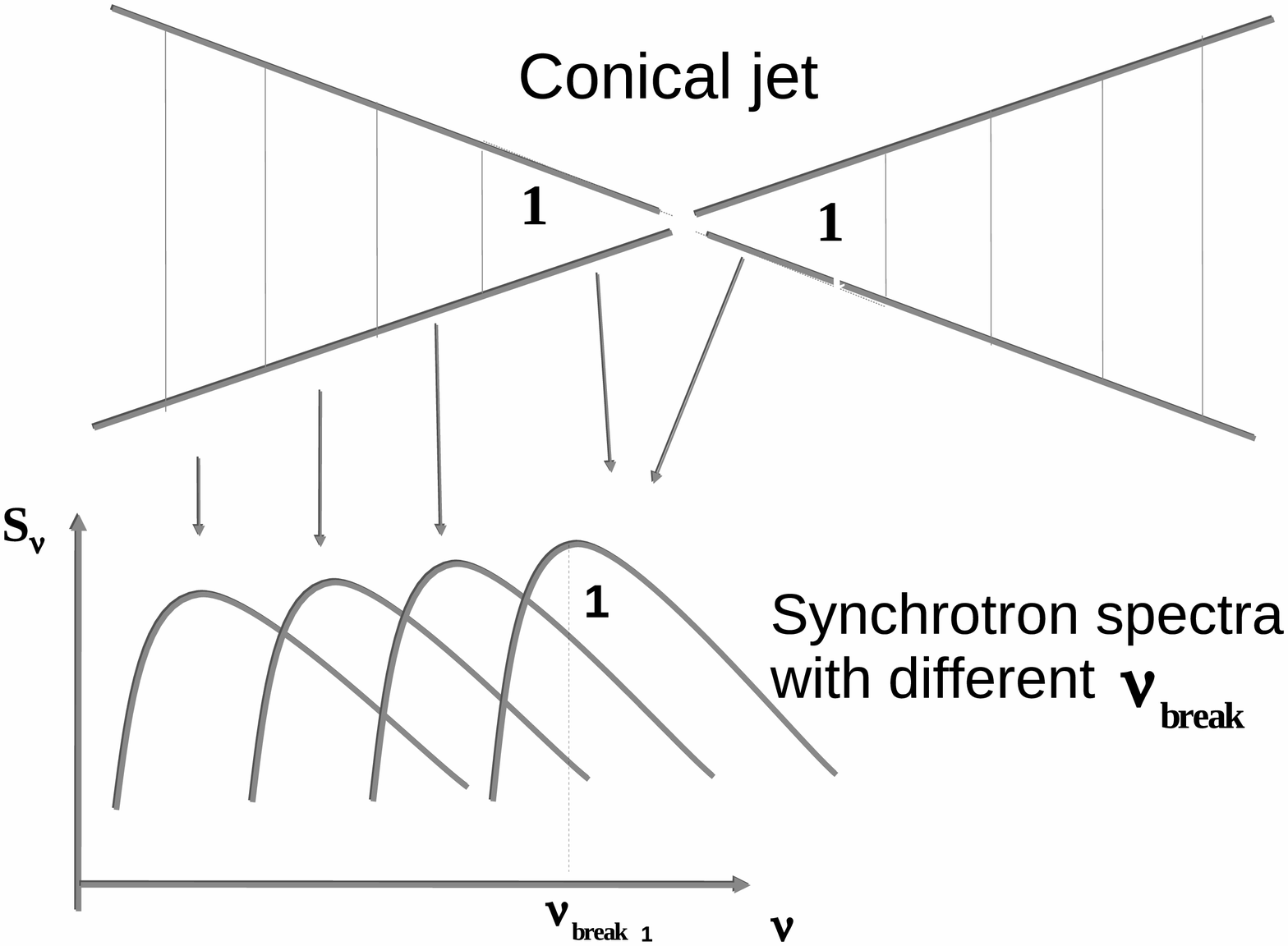}}
\caption{\footnotesize  Composite flat  spectrum as a result of 
superposition of individual spectra
associated at different  jet segments, 
each with a different $\nu_{break}$.}
%%%%%%%appear flat FIGURA CASELLA??}. 
\end{figure}
%
%The power law of the steep power law state extends infact to high energy. 
\begin{figure}
\resizebox{\hsize}{!}{\includegraphics[clip=true]{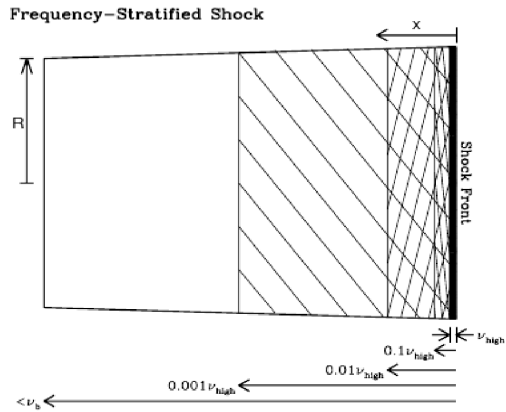}}
\caption{\footnotesize Frequency distribution behind the shock front 
\citep{marscher10}} 
\end{figure}

\section{Discussion}
I comment here on three open questions.
%\begin{enumerate}

%\item 

$1.~~$The transient jet is associated to shocks produced by differences in flow speed.
For the internal shock scenario to be  working 
and giving rise to the bright optically thin radio flare (associated to shocks),
particles (i.e. a new population) must be travelling  with a  higher velocity with
respect to
the pre-existing low/hard state continuous flow. 
What kind of process can generate these fast particles?
Moreover, a related open issue is the puzzling rather short timescale involved.
The low/ hard phase is a relatively long and stable phase, lasting 
tens of days. 
\citet{belloni05}  analysed in X-rays in great detail 
 the point of the transition from the low/hard state,  
distinguishing between before and after the jet line (i.e. when the transient jet is
generated) 
into two additional states - called hard intermediate state (HIMS)  and soft intermediate
state (SIMS). They noticed, how sharp the transition can be, sometimes over only a few seconds
\citep{motta09}.
One kind of energy that can be built up and accumulated 
over long time scales  and then dissipated over very short time
scales clearly is  magnetic energy, as we see in solar flares \citep{ komissarov07}.
%(Komissarov et al. 2006).
Furthermore, also as in solar flares, magnetic reconnection can accelerate particles 
to relativistic velocities.
%These particles  travelling in the steady flow would create shocks and therefore
%the optically thin outburst.
Therefore, magnetic reconnection would be, on the basis of these two arguments, timescales and production of fast particles, a quite  good candidate for triggering the transition. 
%the physical mechanism at the basis of the particle acceleration.
%<a href="http://arxiv.org/PS_cache/astro-ph/pdf/0606/0606375v2.pdf" %target="_blank">http://arxiv.org/PS_cache/astro-ph/pdf/0606/0606375v2.pdf</a>
As a matter of fact, the prolonged  removal of angular momentum from the accretion
disk via the steady jet
has a very important  effect
on the overall process of the accretion process.
As  proved for the bipolar outflows from young stellar objects,
the jets are capable of extracting two thirds or more of the excess angular momentum
present in the disk \citep{woitas05}.
This loss of  angular momentum slows down the disk material to sub-Keplerian
rotation and therefore the
disk matter can finally accrete onto the central object \citep{matsumoto96}.
%(Matsumoto et al. 1996).
This increase in accreted matter onto the compact object implies that the material pulls
the deformed magnetic field with it even further. The magnetic field
compression is thus increased and magnetic reconnection may occur 
\citep{novikov73, matsumoto96, gouveia05}.
% (Novikov \& Thorne 1973, Matsumoto et al. 1996, Gouveia dal Pino 2005).
In addition,
as observed in the Sun, energy released during magnetic reconnection goes into
electron acceleration, but in some cases also
into bulk motion,
%Magnetic recoonection can accelerate particles in random motion but can
%also goes in bulk motion, 
propelling a flux rope away, as it occurs in a coronal mass ejection (CME)
\citep{qiu04}.
\citet{yuan09}, in analogy with
the solar CMEs, propose a magnetohydrodynamical model for transient  jets,
where the readjustment of the magnetic field provides the free energy to drive CME-like ejections of plasmoids. 
\citet{yuan09} suggest that the interaction of  
plasmoids, ejected at very high-speed, with the
slow preexisting continuous jet  could lead to a shock formation.
%The interesting aspect of the study of  \citet{yuan09} is that this  model would 
%not only explain the optically thin outburst, associated with the interaction between a
%CME-like plasmoid and the steady jet, but also the presence of a preceding optically
%thick outburst, associated to the CME-like plasmoid itself.   
Following \citet{yuan09} the  expansion of the plasmoid
would create  delayed peaks in the light curves at different wavelengths
as indeed observed 
in  \lsi (Fig. 4) and 
in  GRS 1915+105 by \citet{mirabel98}.
In addition, radio emission associated to some CME has been succesfully
modelled as synchrotron radiation.
\citet{bastian07},
%Bastian 2007 
analizing a radio burst associated to a  CME, 
reproduced the drift to lower frequencies with time of the flux maximum 
by assuming a decrease  of magnetic field and particle density due to 
source expansion.
%Finally, a CME like ejection could occur  in only one side 
% of the accretion disk,  resulting
%in an asymmetric ejection. 
%This could  explain observations   
%%%the never understood observations of 
%like those of GRO J1655-40  by \citet{hjellming95}  
%where the receding jet sometimes appears brighter than the approaching one.

$2.~~$
Observations of \grs by \citet{mirabelrodriguez99} (their Fig. 2, especially the map
of April 30)
and by \citet{fender99} (their Fig. 2, especially epoch 736.7), 
showing  moving components/shocks,   seem to lack emission from the center.
Is the core destroyed by the transient?
The lack of a standing shock  could find an explanation in the   
changed conditions during the first phase of the shock and probably of broken
symmetry of the conical jet.
In the AGN NRAO~150 there is  a sequence of observations at 86 GHZ and at 43 GHz
by  \citet{agudo07} (their Figure 3) showing that 
associated to the onset of fast  moving  components  
there are  pronounced dips in the core light curve.
At one epoch, 2006.22, rather than only a dip there seems to be a real lack of 
the core; this lack is confirmed also by the presented sequence of images. 
\citet{agudo07} suggest  a  possible change of Doppler boosting factors to explain
this puzzling dip. 
However, the comparison with the microquasar \grs  might suggest 
the alternative hypothesis 
that the dip could be a real one and that the core is, for a short interval,
missing.

$3.~~ $The above comparison between  \grs and  NRAO~150 shows how
important it would be to extrapolate the knowledge on steady and transient jets in 
microquasars to AGNs.
A comparison between the two kinds of classes presents obvious difficulties.
Jets in microquasars are of the order of hundreds of AU: the relationship
between jet variations and activity in the core is streightforward. In AGNs 
the distant  jet components may be completly detached from the present activity of
the core.
In AGNs the steady jet  remains visible under/around the shocked regions:
one observes a  sequence of bright and typically optically thin regions that move 
at superluminal speeds, embedded in the steady jet, which is in the AGN community called 
``underlying flow''.
Still, besides these difficulties, there exist several interesting parallels among the
two classes of objects
\citep{kording06,uttley06,camenzind08}.
Radio quiet quasars would  correspond to the thermal state (Fig. 5),
BL LAcs and FR-I radio galaxies  to the low/hard state and finally 
the transient jet, or steep power-law state would correspond  to
the FR-II radio galaxies.

\begin{acknowledgements}
It is a pleasure to  acknowledge the
helpful discussions with Manolis Angelakis, 
%Tim Bastian, 
Christian Fendt,  Svetlana Jorstad, Sergei Komissarov, 
Andrei Lobanov, J\"urgen Neidh\"ofer, Giannina Poletto, Lisa Zimmermann and  Francesca Zuccarello. 
The Green Bank Interferometer is a facility of the National
Science Foundation operated by the NRAO in support of
NASA High Energy Astrophysics programs.
\end{acknowledgements}

\bibliographystyle{aa}

\end{document}